# Expected geoneutrino signal at JUNO

Virginia STRATI[1,2,*], Marica BALDONCINI[1,3], Ivan CALLEGARI[2], Fabio MANTOVANI[1,3],

William F. McDONOUGH[4], Barbara RICCI[1,3], Gerti XHIXHA[2]

[1] Department of Physics and Earth Sciences, University of Ferrara, Via Saragat 1, 44121 - Ferrara, Italy

[2] INFN, Legnaro National Laboratories, Viale dell'Università, 2 - 35020 Legnaro (Padua) Italy

[3] INFN, Ferrara Section, Via Saragat 1, 44121 - Ferrara, Italy

[4] Department of Geology, University of Maryland, 237 Regents Drive, College Park, MD 20742, USA.

*Corresponding author Email: strati@fe.infn.it

**Abstract**

Constraints on the Earth's composition and on its radiogenic energy budget come from the detection of geoneutrinos. The KamLAND and Borexino experiments recently reported the geoneutrino flux, which reflects the amount and distribution of U and Th inside the Earth. The JUNO neutrino experiment, designed as a 20 kton liquid scintillator detector, will be built in an underground laboratory in South China about 53 km from the Yangjiang and Taishan nuclear power plants, each one having a planned thermal power of approximately 18 GW. Given the large detector mass and the intense reactor antineutrino flux, JUNO aims to collect high statistics antineutrino signals from reactors but also to address the challenge of discriminating the geoneutrino signal from the reactor background.

The predicted geoneutrino signal at JUNO is $39.7^{+6.5}_{-5.2}$ TNU, based on the existing reference Earth model, with the dominant source of uncertainty coming from the modeling of the compositional variability in the local upper crust that surrounds (out to ~500 km) the detector. A special focus is dedicated to the 6° × 4° Local Crust surrounding the detector which is estimated to contribute for the 44% of the signal. On the base of a worldwide reference model for reactor antineutrinos, the ratio between reactor antineutrino and geoneutrino signals in the geoneutrino energy window is estimated to be 0.7 considering reactors operating in year 2013 and reaches a value of 8.9 by adding the contribution of the future nuclear power plants.

In order to extract useful information about the mantle's composition, a refinement of the abundance and distribution of U and Th in the Local Crust is required, with particular attention to the geochemical characterization of the accessible upper crust where 47% of the expected geoneutrino signal originates and this region contributes the major source of uncertainty.

**Background**

The first experimental evidence of geoneutrinos, i.e. electron antineutrinos produced in beta decays along the $^{238}$U and $^{232}$Th decay chains, was claimed by the KamLAND Collaboration in 2005 (KamLAND Collaboration 2005), which ushered in a new method for exploring the Earth's interior and provided constraints on the planet's composition and specifically its radiogenic element budget (Fiorentini et al. 2007). The geoneutrino energy spectrum contains in it distinctive contributions from U and Th, each one resulting from different rates and shapes of their decays (see Figure 3 and Figure 5 of (Fiorentini et al. 2007)) and from concentrations and spatial distributions of these elements inside the Earth.

Geoneutrinos are measured in liquid scintillation detectors via the Inverse Beta Decay (IBD) reaction on free protons:

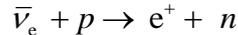

$$\bar{\nu}_e + p \rightarrow e^+ + n$$

whose energy threshold of 1.806 MeV means that only a small fraction of the antineutrinos produced from the U and Th decay chains are detectable. The IBD detection event in liquid scintillator produces two flashes of light: the annihilation flash, from electron-positron interaction, followed by the deuterium formation flash, which is 2.2 MeV of light that follows some 200 μs later. The delayed coincidence of these two flashes of light provides the critical identification of the antineutrino interaction and eliminates most background events. The KamLAND and Borexino experiments recently reported $116^{+28}_{-27}$ geoneutrino events over 2991 days (KamLAND Collaboration 2013) and 14.3 ± 4.4 geoneutrino events in 1353 days (Borexino Collaboration 2013), respectively. Differences in the detection rates reflect the detector sizes, with the KamLAND detector being ~1kton and the Borexino detector 0.3 kton.

The most significant source of background for geoneutrino measurements is due to reactor antineutrinos, i.e. electron antineutrinos emitted during the beta decays of fission products from $^{235}$U, $^{238}$U, $^{239}$Pu and $^{241}$Pu burning. Approximately 30% of the reactor antineutrino events are recorded in the geoneutrino energy window extending from the IBD threshold up to the endpoint of the $^{214}$Bi beta decay spectrum (3.272 MeV) (Fiorentini et al. 2010). The Terrestrial Neutrino Unit (TNU), which is the signal that corresponds to one IBD event per $10^{32}$ free protons per year at 100% efficiency, is used to compare the different integrated spectral components (i.e. antineutrinos from U, Th and reactors) measured by the detectors or just beneath the Earth's surface.

In the past decade reactor antineutrino experiments played a decisive role in unraveling the neutrino puzzle, which currently recognizes three flavor eigenstates ($v_e$, $v_\mu$, $v_\tau$), each of which mixes with three mass eigenstates ($v_1$, $v_2$, $v_3$) via three mixing angles ($\theta_{12}$, $\theta_{13}$, $\theta_{23}$). The quantities that govern the oscillation frequencies are two differences between squared masses, (i.e. $\delta m^2 = m_2^2 - m_1^2 > 0$ and $\Delta m^2 = m_3^2 - (m_1^2 + m_2^2)/2$). Central to neutrino studies is understanding the neutrino mass hierarchy (i.e. $\Delta m^2 > 0$ or $\Delta m^2 < 0$) (Capozzi et al. 2014; Ge et al. 2013).

Massive (>10kton) detectors such as the JUNO (Li 2014) and Reno-50 (Kim 2013) experiments are being constructed at medium baseline distances (a few tens of km) away from bright reactor antineutrino fluxes in order to assess significant physics goals regarding the neutrino properties, in first place the mass hierarchy. These experiments intend also to obtain sub-percent precision measurements of neutrino oscillation parameters and along the way make observations of events of astrophysical and terrestrial origin.

The Jiangmen Underground Neutrino Observatory (JUNO) is located (N 22.12° E 112.52°) in Kaiping, Jiangmen, Guangdong province (South China), about 53 km away from the Yangjiang and Taishan nuclear power plants, which are presently under construction. The combined thermal power of these two units is planned to be on the order of 36 GW (Li and Zhou 2014) (Figure 1). The JUNO experiment is designed as a liquid scintillator detector of 20 kton mass that will be built in a laboratory some 700 m underground (approximately 2000 m water equivalent). This amount of overburden will attenuate the cosmic muon flux, which contributes to the overall detector background signal, but this overburden is significantly less than that at the KamLAND and Borexino experiments. The detector energy response and the spatial distribution of the reactor cores are the most critical features affecting the experimental sensitivity (Li et al. 2013) required to achieve the intended physics goals.

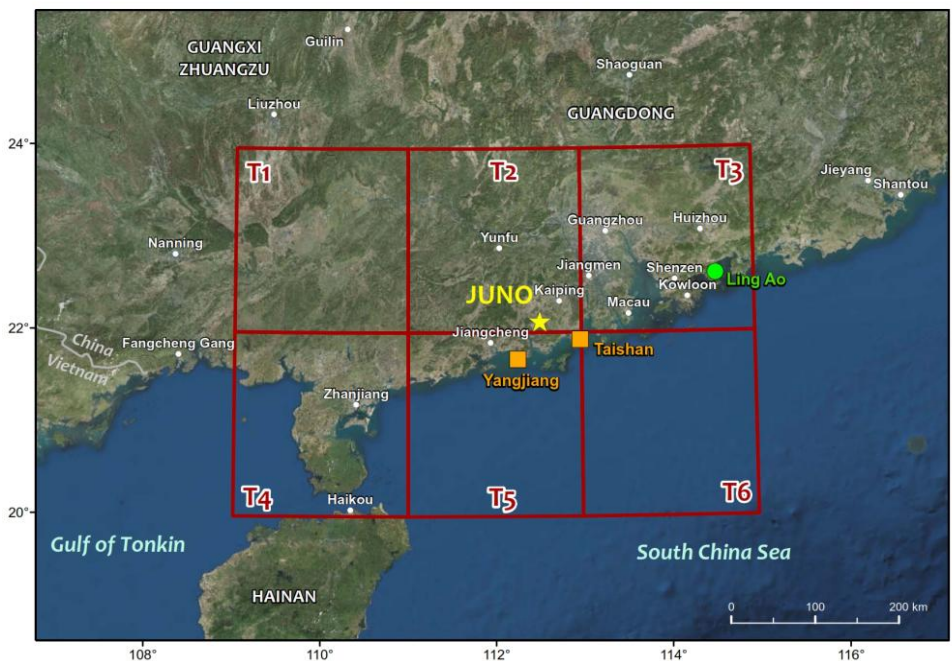

**Figure 1.** Map of LOC surrounding JUNO. JUNO (yellow star) is located in Kaiping, Jiangmen, Guangdong province (South China) and the planned (orange square) and operational (green circle) nuclear power plants. The six $2° \times 2°$ Tiles (dark red lines) define the LOC.

The goal of this present study is to predict the geoneutrino signal at JUNO on the basis of an existing reference Earth model (Huang et al. 2013), together with an estimate of the expected reactor antineutrino signal. Since a significant contribution to the expected geoneutrino signal comes from U and Th in the continental crust surrounding the site, we follow past approaches to study the local contribution (Coltorti et al. 2011; Fiorentini et al. 2012; Huang et al. 2013; Huang et al. 2014), with a particular interest in focusing in on the closest 6° × 4° grid surrounding the detector, we define this latter region as the LOcal Crust (LOC) (Figure 1).

**Methods**

The geoneutrino signal expected at JUNO is calculated adopting the same methodology and the same inputs of the reference Earth model developed in (Huang et al. 2013). It provides a description of the abundances and distribution of the Heat-Producing Elements (HPEs, i.e. U, Th and K) in the Earth's crust, along with their uncertainties. According to this model the silicate portion of the Earth is composed of five dominant reservoirs: the Depleted Mantle (DM), the Enriched Mantle (EM), the Lithospheric Mantle (LM), the Continental Crust (CC) and the Oceanic Crust (OC). The continental crust is dominantly composed of Lower Crust (LC), Middle Crust (MC) and Upper Crust (UC) and it is overlain by shallow layers of Sediments (Sed) which also covers the OC.

The surface geoneutrino flux is calculated by dividing the Earth surface in 1° × 1° tiles that are projected vertically into discrete volume cells and each cell is assigned physical and chemical states. Just for the sake of computing flux, the 1° × 1° tiles are further subdivided into many subcells with the same properties of the parent tile. The number of subcells is progressively bigger approaching the detector location with the aim of not introducing any bias due to discretization.

The total crustal thickness of each cell and its associated uncertainty correspond, respectively, to the mean and the half-range of three crustal models obtained from different approaches: the global crustal model based on reflection and refraction data "CRUST 2.0" (Bassin et al. 2000; Laske et al. 2001), the global shear-velocity model of the crust and upper mantle "CUB 2.0" (Shapiro and Ritzwoller 2002) and the high-resolution map of Moho (crust-mantle boundary) depth based on gravity field data "GEMMA" (Reguzzoni and Tselfes 2009; Reguzzoni and Sampietro 2015).The reference model incorporates the relative proportional thickness of the crustal layers along with density and elastic properties (compressional and shear waves velocity) reported in CRUST 2.0. The same information are adopted for the Sed layer using the global sediment map of (Laske and Masters 1997). In Figure 2 the thicknesses of the continental crust layers in the 24 cells constituting the LOC for JUNO are reported. Their total crustal thickness ranges between 26.3 km and 32.3 km with an uncertainty for each cell of ~7%.

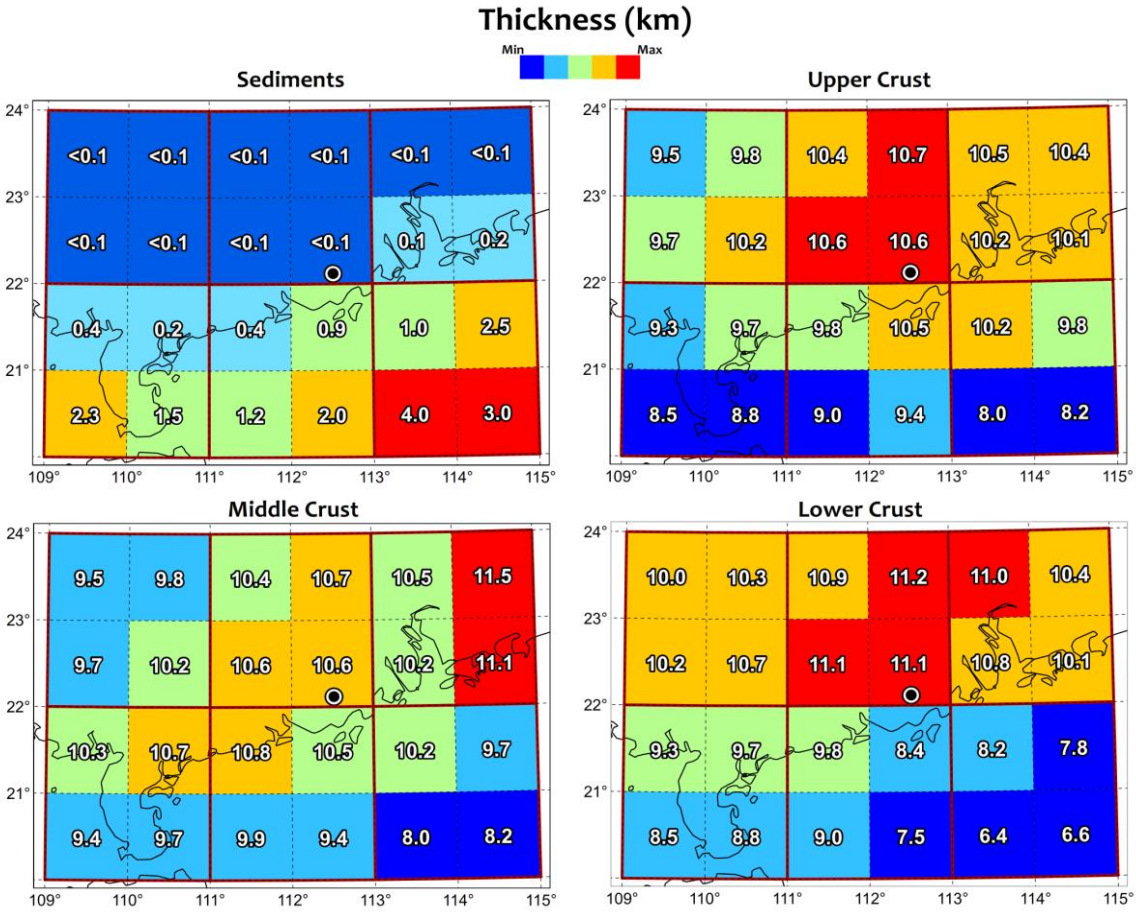

**Figure 2.** Thicknesses of the 4 crustal layers in the LOC. The thicknesses in km of the Sed, UC, MC and LC layers are reported for each of the 24 cells constituting the LOC surrounding JUNO (black circle), with color coding to illustrate gradients in thickness.

The HPEs abundances in the Sed, OC and UC layers are assumed to be relatively homogenous and correspond to the values reported in Table 3 of (Huang et al. 2013). The ratio between Felsic and Mafic components in the deep CC (MC and LC) is inferred from seismic velocity data and these data are in turn used to estimate the U and Th content of each cell of the reference crustal model. Focusing on the LOC, the central values of U abundance in MC and LC vary in the range 0.8 - 1.2 µg/g and 0.3 - 0.1 µg/g, respectively. The Th/U ratio in deep CC of the LOC is typically ~5 as compared to a bulk silicate Earth ratio of 3.9 or a bulk CC ratio just greater than 4.0; the higher Th/U ratio in the deep CC is likely due to the greater upward mobility of U during dehydration reactions that accompany granulite facies metamorphism of the deep CC.

In the reference model of (Huang et al. 2013) the Lithospheric Mantle (LM) corresponds to the portion of Earth between the Moho discontinuity and an assumed standard depth of 175 km beneath the surface. The thickness of this unit in the LOC ranges between 143 km and 149 km and its composition is modeled from the database reported in (McDonough 1990) and the update in (Huang et al. 2013). In our calculation we adopt for the LM the U and Th abundances of $0.03^{+0.05}_{-0.02}$ µg/g and $0.15^{+0.28}_{-0.10}$ µg/g, respectively (Huang et al. 2013).

The sublithospheric mantle extends down from the base of the lithosphere to the core-mantle boundary and is divided in two spherically symmetric domains, the Depleted Mantle (DM) and the Enriched Mantle (EM), whose density profiles are derived from the Preliminary Reference Earth Model, "PREM" (Dziewonski and Anderson 1981). Adopting a mass ratio $M_{DM} / M_{EM} = 4.56$ (Huang et al. 2013), we calculate the masses of these two reservoirs $M_{DM} = 3.207 \cdot 10^{24}$ kg and $M_{EM} = 0.704 \cdot 10^{24}$ kg. In a survey of MidOcean Ridge Basalts (MORB) (Arevalo and McDonough 2010) reported the log-normal based average abundances of uranium ($U_{MORB}$ = 80 ng/g) and thorium ($Th_{MORB}$ = 220 ng/g), and from this calculated the $U_{DM}$ = 8 ng/g and $Th_{DM}$ = 22 ng/g based on a simple melting model. Based on these assumptions the $U_{EM}$ can be calculated:

$$U_{EM} = \frac{m_{BSE} - m_C}{M_{EM}} - U_{DM}\frac{M_{DM}}{M_{EM}}$$

where $m_{BSE}$ = 8.1 ·$10^{16}$ kg is the U mass in the Bulk Silicate Earth (BSE) (McDonough and Sun 1995) and $m_C$ = 3.1 ·$10^{16}$ kg is the total U mass in the crust (Huang et al. 2013). The mantle geoneutrino signals reported in Table 1 are calculated with $U_{DM}$ = 8 ng/g and $U_{EM}$ = 34 ng/g together with $(Th/U)_{DM}$ = 2.8 and $(Th/U)_{EM}$ = 4.8.

**Table 1. Geoneutrino signals from U and Th expected in JUNO.** The inputs for the calculations are taken fromn (Huang et al. 2013) and the signals from different reservoirs indicated in the first column are in TNU.

|  | S (U) | S (Th) | S (U+Th) |
|---|---|---|---|
| Sed CC | $0.5^{+0.1}_{-0.1}$ | $0.16^{+0.02}_{-0.02}$ | $0.64^{+0.1}_{-0.1}$ |
| UC | $14.6^{+3.5}_{-3.4}$ | $3.9^{+0.5}_{-0.5}$ | $18.5^{+3.6}_{-3.4}$ |
| MC | $4.7^{+3.0}_{-1.8}$ | $1.7^{+1.6}_{-0.8}$ | $6.8^{+3.6}_{-2.3}$ |
| LC | $0.9^{+0.7}_{-0.4}$ | $0.4^{+0.7}_{-0.2}$ | $1.5^{+1.0}_{-0.6}$ |
| Sed OC | $0.08^{+0.02}_{-0.02}$ | $0.03^{+0.01}_{-0.01}$ | $0.11^{+0.02}_{-0.02}$ |
| OC | $0.05^{+0.02}_{-0.02}$ | $0.01^{+0.01}_{-0.01}$ | $0.06^{+0.02}_{-0.02}$ |
| Bulk Crust | $21.3^{+4.8}_{-4.2}$ | $6.6^{+1.9}_{-1.2}$ | $28.2^{+5.2}_{-4.5}$ |
| CLM | $1.3^{+2.4}_{-0.9}$ | $0.4^{+1.0}_{-0.3}$ | $2.1^{+2.9}_{-1.3}$ |
| Total Lithosphere | $23.2^{+5.9}_{-4.8}$ | $7.3^{+2.4}_{-1.5}$ | $30.9^{+6.5}_{-5.2}$ |
| DM | 4.2 | 0.8 | 4.9 |
| EM | 2.9 | 0.9 | 3.8 |
| **Gran Total** | $30.3^{+5.9}_{-4.8}$ | $9.0^{+2.4}_{-1.5}$ | $39.7^{+6.5}_{-5.2}$ |

## Results and discussion

In Table 1 we summarize the contributions to the expected geoneutrino signal at JUNO produced by U and Th in each of the reservoirs identified in the model. The central value and the asymmetric uncertainties are respectively the median and 1σ errors of a positively skewed distribution obtained from Monte Carlo simulation. This approach was developed for the first time in (Huang et al. 2013) in order to combine Gaussian probability density function of geophysical and (some) geochemical inputs, together with the lognormal distributions of U and Th abundances observed in the felsic and mafic rocks of MC and LC.

The total geoneutrino signal at JUNO is $G = 39.7^{+6.5}_{-5.2}$ TNU where the 1σ error only recognizes the uncertainties of the inputs of the lithosphere, which are mainly due to the uncertainties in the composition of the rocks and subsequently to the geophysical inputs. The predicted mantle contribution at JUNO is assumed to be $S_M \approx 9$ TNU known (Huang et al. 2013). The expected geoneutrino signal from the mantle is essentially model dependent and it is estimated according to a mass balance argument. Uncertainty in the assumed mantle model is much less than that predicted for the lithosphere (e.g. δG ≈ ±6 TNU). An extensive discussion of different mantle's structure is described in (Šrámek et al. 2013), which considers a range of geoneutrino signals for different mantle's models.

Thus, a future refinement of the abundances and distribution of HPEs in the UC surrounding the JUNO detector is strongly recommended, as this region provides ~ 47% of G and is a significant contributor to the total uncertainty.

Plotting the cumulative geoneutrino signal as a function of the distance from JUNO for the different Earth reservoirs (Figure 3), we observe that half of the total signal comes from U and Th in the regional crust that lies within 550 km of the detector. Since the modeling of the geoneutrino flux is based on 1° × 1° cells, we study the signal produced in LOC subdivided in six 2°× 2° tiles (Figure 1).

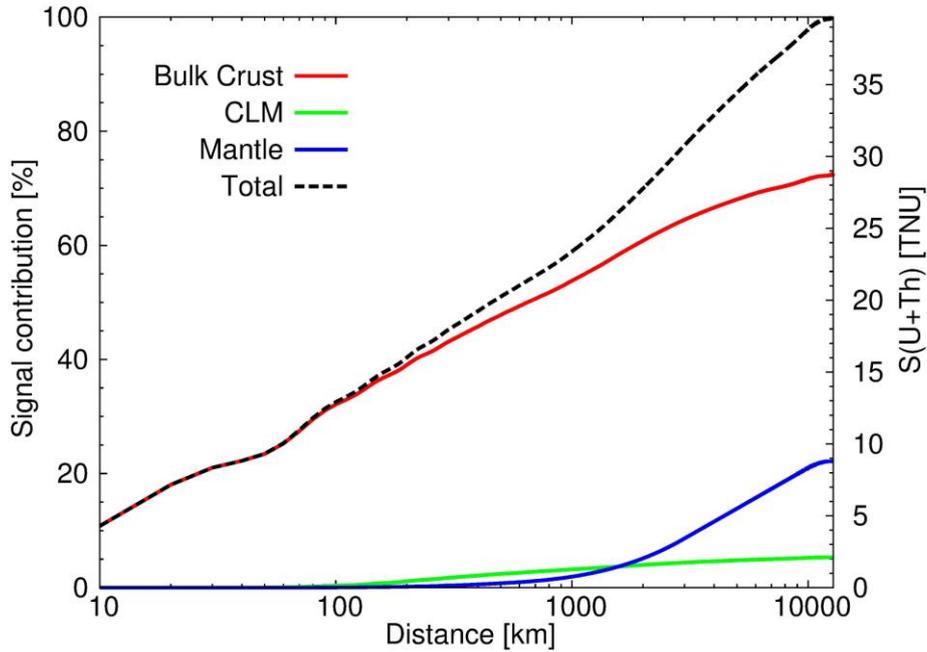

**Figure 3.** Geoneutrino signal contribution. The cumulative geoneutrino signal and the percentage contributions of the Bulk Crust, Continental Lithospheric Mantle (CLM) and Mantle are represented as function of the distance from JUNO.

The geoneutrino signals from U and Th in the lithosphere of each tile are reported in Table 2 with their uncertainties. The main contribution (27% of G) comes from tile T2 in which the JUNO experiment is located (Figure 1). The thick UC in this tile, which is covered by a very shallow layer of Sed (Figure 2), is predicted to give a signal of $7.6^{+1.5}_{-1.4}$ TNU. Therefore a refined study of the U and Th content of the UC in tile T2 is a high-value target for improving the accuracy and precision of the predicted geoneutrino signal at JUNO. Evaluating the antineutrino signal requires knowledge of several ingredients necessary for modeling the three antineutrino life stages: production, propagation to the detector site and detection in liquid scintillation detectors via the IBD reaction. The propagation and detection processes are independent by the source of the particles and we modeled these two stages using the oscillation parameters from (Ge et al. 2013) and the IBD cross section from (Strumia and Vissani 2003). Spectral parameters for U and Th geoneutrinos are from (Fiorentini et al. 2007) and modulation of these fluxes are based on (Huang et al. 2013). Reactor antineutrino production is calculated adopting data from a worldwide reference model from (Baldoncini et al. 2014). Reported in Figure 4 are the energy distributions of geoneutrinos and reactor antineutrino signals in two different scenarios: in the full energy region $R_{OFF} = 95.3^{+2.6}_{-2.4}$ TNU is obtained with data from worldwide commercial reactors operating in 2013 and $R_{ON} = 1566^{+111}_{-100}$ TNU, including the Yangjiang (17.4 GW) and Taishan (18.4 GW) nuclear power plants operating at a 80% annual average load factor (Baldoncini et al. 2014). In the geoneutrino energy window (i.e. 1.806 - 3.272 MeV) the reactor signal is $S_{OFF} = 26.0^{+2.2}_{-2.3}$ TNU and $S_{ON} = 354^{+45}_{-41}$ TNU (Table 3). Assuming a scenario whereby JUNO's signal does not have a background signal from Yangjiang and Taishan nuclear power plants, the ratio of $S_{OFF}/G = 0.7$, which compares to a value of 0.6 for the Borexino detector (Baldoncini et al. 2014). Considering only the statistical uncertainties, in the $R_{OFF}$ scenario JUNO is an excellent experiment for geoneutrino measurements reaching a 10% accuracy on the geoneutrino signal in approximately 105 days (assuming a $C_{17}H_{28}$ liquid scintillator composition, a 100% detection efficiency and reactor antineutrinos as the sole source of background), given 576 geoneutrino events per year for a target mass of $14.5 \cdot 10^{32}$ free protons. This optimistic expectation doesn't take into account the uncertainties of $S_{OFF}$ and the background due to production of cosmic-muons spallation, accidental coincidences and radioactive contaminants in the detector.

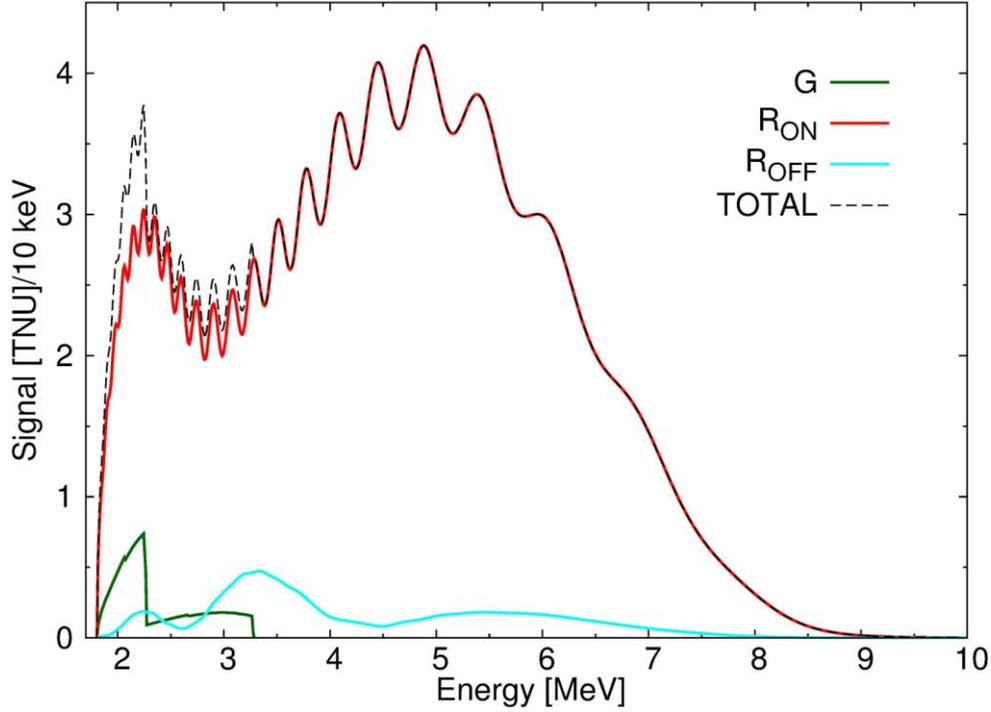

**Figure 4.** Antineutrino energy spectra expected at JUNO. Geoneutrino energy spectrum (green) is reported together with the energy reactor antineutrino spectra computed considering the commercial reactors operating all over the world in 2013 (cyan) and adding the contribution of the Yangjiang and Taishan nuclear power plants (red). The reactor antineutrino spectra are computed assuming normal hierarchy and neutrino oscillation. The total spectrum (black dashed lines) is obtained assuming the $R_{ON}$ scenario.

**Table 2. Geoneutrino signals from 6 tiles of the LOC.** The expected geoneutrino signal in TNU from U and Th contained in the lithosphere (CC+CLM) of the 6 tiles in Figure 1. In the last column contributions in percentage are reported.

| Tile | S (U) | S (Th) | S (U+Th) | Percentage |
|------|-------|--------|----------|------------|
| T1 | $0.4^{+0.1}_{-0.1}$ | $0.1^{+0.1}_{-0.1}$ | $0.5^{+0.1}_{-0.1}$ | 3.0 |
| T2 | $8.1^{+1.9}_{-1.7}$ | $2.6^{+0.8}_{-0.5}$ | $10.8^{+2.1}_{-1.8}$ | 62.1 |
| T3 | $1.1^{+0.3}_{-0.2}$ | $0.4^{+0.2}_{-0.1}$ | $1.5^{+0.3}_{-0.3}$ | 8.6 |
| T4 | $0.3^{+0.1}_{-0.1}$ | $0.1^{+0.1}_{-0.1}$ | $0.4^{+0.1}_{-0.1}$ | 2.2 |
| T5 | $2.5^{+0.5}_{-0.5}$ | $0.7^{+0.2}_{-0.1}$ | $3.2^{+0.6}_{-0.5}$ | 18.2 |
| T6 | $0.8^{+0.2}_{-0.2}$ | $0.2^{+0.1}_{-0.1}$ | $1.0^{+0.2}_{-0.2}$ | 5.9 |

**Table 3. Geoneutrino and reactor antineutrinos signals at JUNO.** The Gran total geoneutrino signal (G) is the sum of the Local contribution ($S_{LOC}$), from the rest of the crust, i.e. Far Field Crust, ($S_{FFC}$) and from the Mantle ($S_M$). The reactor antineutrino signal in the geoneutrino window is calculated from data referred for commercial reactors operating all over the world in 2013 ($S_{OFF}$) and adding the contribution of the Yangjiang (17.4 GW) and Taishan (18.4 GW) nuclear power plants ($S_{ON}$) (Baldoncini et al. 2014). All the signals are expressed in TNU.

|  | S [TNU] |
|---|---|
| Local contribution | $17.4^{+3.3}_{-2.8}$ |
| Far Field Crust | $13.4^{+3.3}_{-2.4}$ |
| Mantle | 8.8 |
| **Gran total geoneutrinos** | $39.7^{+6.5}_{-5.2}$ |
| **Reactors OFF** | $26.0^{+2.2}_{-2.3}$ |
| **Reactors ON** | $354^{+45}_{-41}$ |

**Conclusions**

Designed as a 20kton liquid scintillator detector, the JUNO experiment will collect high statistics for antineutrino signals from reactors and form the Earth. In this study we focused on predicting the geoneutrino signal using the Earth reference model of (Huang et al. 2013). The contribution originating from naturally occurring U and Th in the 6°×4° LOcal Crust (LOC) surrounding the JUNO detector (Figure 1) was determined. The main results of this study are summarized as follows.

- The thickness of the Sed, UC, MC and LC layers of the 24 1°×1° cells of the LOC are reported (Figure 2). The Moho depth of the continental LOC ranges between 26.3 km and 32.3 km and the uncertainty for each 1°×1° cell is of the order of 7%.

- The total and local geoneutrino signals at JUNO are $G = 39.7^{+6.5}_{-5.2}$ TNU and $S_{LOC} = 17.4^{+3.3}_{-2.8}$ TNU, respectively. The asymmetric 1σ errors are obtained from Monte Carlo simulations and account only for uncertainties from the lithosphere. The major source of uncertainty comes from predicting the abundances and distribution of U and Th in local crustal rocks.

- High-resolution seismic data acquired in the LOC can improve the present geophysical model of the crust and CLM, of which the latter is assumed to have a homogenous depth of 175 ± 75 km. The CLM composition is derived from data for U and Th abundances inferred from the peridotite xenoliths and its geoneutrino signal is of $2.1^{+2.9}_{-1.3}$ TNU.

- The HPEs in the regional crust extending out to 550 km from the detector produce half of the total expected geoneutrino signal (Figure 3). The U and Th in the 2° × 2° tile that hosts JUNO produces $10.8^{+2.1}_{-1.8}$ TNU corresponding to 27% of G. Since this region is characterized by a thick UC, which gives $7.6^{+1.5}_{-1.4}$ TNU, a refined geophysical and geochemical model of the UC of this tile is highly desired.

- The reactor signal in the geoneutrino window assuming two scenarios is $S_{OFF} = 26.0^{+2.2}_{-2.3}$ TNU with the 2013 reactor operational data only and $S_{ON} = 355^{+44}_{-41}$ TNU when the contributions of the Yangjiang and Taishan nuclear power plants are added. There is a potential to achieve up to 10% accuracy on geoneutrinos after 105 days of data accumulation, under conditions of Yangjiang and Taishan nuclear power plants being off.

The JUNO experiment has the potential to reach a milestone in geoneutrino science, although some technical challenges must be addressed to minimize background (e.g. production of cosmic-muons spallation, accidental coincidences, radioactive contaminants in the detector). Assuming $S_{OFF}/G = 0.7$, JUNO can collect hundreds of low background geoneutrino events in less than a year under optimal conditions. A future refinement of the U and Th distribution and abundance in the LOC is strongly recommended. Such data will lead to insights on the radiogenic heat production in the Earth, the composition of the mantle and constraints on the chondritic building blocks that made the planet.

## List of abbreviations used

IBD, Inverse Beta Decay;

BSE, Bulk Silicate Earth;

TNU, Terrestrial Neutrino Unit;

JUNO, Jiangmen Underground Neutrino Observatory;

HPEs, Heat Production Elements;

DM, Depleted Mantle;

EM, Enriched Mantle;

LM, Lithospheric Mantle;

CC, Continental Crust;

OC, Oceanic Crust;

Sed, Sediments;

UC, Upper Crust;

MC, Middle Crust;

LC, Lower Crust;

CLM , Continental Lithospheric Mantle.

## Acknowledgements

We are grateful to R. L. Rudnick and Y. Huang for fruitful discussions on crustal modeling of geoneutrino fluxes. We appreciate observations on the geoneutrino signal predictions from S. Dye, G. Fiorentini, L. Ludhova and H. Watanabe. We thank J. Mandula for the valuable help in compiling the nuclear reactor database. We wish to thank two anonymous reviewers for their detailed and thoughtful reviews. This work was partially supported by the Istituto

Nazionale di Fisica Nucleare (INFN) through the ITALRAD Project, by the University of Ferrara through the research initiative "Fondo di Ateneo per la Ricerca scientifica FAR 2014" and partially by the U.S. National Science Foundation Grants EAR 1067983/1068097.

**References**

Arevalo R, McDonough WF (2010) Chemical variations and regional diversity observed in MORB. Chemical Geology 271 (1-2):70-85. doi:10.1016/j.chemgeo.2009.12.013

Baldoncini M, Callegari I, Fiorentini G, Mantovani F, Ricci B, Strati V, Xhixha G (2014) A reference worldwide model for antineutrino from reactors. arXiv:14116475 [physicsins-det]

Bassin C, Laske G, Masters TG (2000) The current limits of resolution for surface wave tomography in North America. EOS Trans

Borexino Collaboration (2013) Measurement of geo-neutrinos from 1353 days of Borexino. Physics Letters B 722 (4-5):295-300. doi:http://dx.doi.org/10.1016/j.physletb.2013.04.030

Capozzi F, Lisi E, Marrone A (2014) Neutrino mass hierarchy and electron neutrino oscillation parameters with one hundred thousand reactor events. Physical Review D 89 (1):013001

Coltorti M, Boraso R, Mantovani F, Morsilli M, Fiorentini G, Riva A, Rusciadelli G, Tassinari R, Tomei C, Di Carlo G, Chubakov V (2011) U and Th content in the Central Apennines continental crust: A contribution to the determination of the geo-neutrinos flux at LNGS. Geochimica et Cosmochimica Acta 75 (9):2271-2294. doi:10.1016/j.gca.2011.01.024

Dziewonski AM, Anderson DL (1981) Preliminary reference Earth model. Physics of the Earth and Planetary Interiors 25:297-356. doi:10.1016/0031-9201(81)90046-7

Fiorentini G, Fogli G, Lisi E, Mantovani F, Rotunno A (2012) Mantle geoneutrinos in KamLAND and Borexino. Physical Review D 86 (3). doi:10.1103/PhysRevD.86.033004

Fiorentini G, Ianni A, Korga G, Lissia M, Mantovani F, Miramonti L, Oberauer L, Obolensky M, Smirnov O, Suvorov Y (2010) Nuclear physics for geo-neutrino studies. Physical Review C 81 (3). doi:10.1103/PhysRevC.81.034602

Fiorentini G, Lissia M, Mantovani F (2007) Geo-neutrinos and earth's interior. Physics Reports 453

(5-6):117-172. doi:10.1016/j.physrep.2007.09.001

Ge S-F, Hagiwara K, Okamura N, Takaesu Y (2013) Determination of mass hierarchy with medium baseline reactor neutrino experiments. J High Energ Phys 2013 (5):1-23. doi:10.1007/JHEP05(2013)131

Huang Y, Chubakov V, Mantovani F, Rudnick RL, McDonough WF (2013) A reference Earth model for the heat-producing elements and associated geoneutrino flux. Geochemistry, Geophysics, Geosystems 14 (6):2023-2029. doi:10.1002/ggge.20129

Huang Y, Strati V, Mantovani F, Shirey SB, McDonough WF (2014) Regional study of the Archean to Proterozoic crust at the Sudbury Neutrino Observatory (SNO+), Ontario: Predicting the geoneutrino flux. Geochemistry, Geophysics, Geosystems 15 (10):3925 - 3944. doi:10.1002/2014gc005397

KamLAND Collaboration (2005) Measurement of Neutrino Oscillation with KamLAND: Evidence of Spectral Distortion. Physical Review Letters 94 (8):081801

KamLAND Collaboration (2013) Reactor on-off antineutrino measurement with KamLAND. Physical Review D 88 (3):033001

Kim S-B Proposal for RENO-50. In: International Workshop on RENO-50, Seoul National University, Korea, June 13-14, 2013.

Laske G, Masters TG (1997) A global digital map of sediment thickness. EOS Trans AGU 78 F483

Laske G, Masters TG, Reif C (2001) CRUST 2.0: A new global crustal model at 2 x 2 degrees.

Li Y-F, Cao J, Wang Y, Zhan L (2013) Unambiguous determination of the neutrino mass hierarchy using reactor neutrinos. Physical Review D 88 (1):013008

Li Y-F, Zhou Y-L (2014) Shifts of neutrino oscillation parameters in reactor antineutrino experiments with non-standard interactions. Nuclear Physics B 888 (0):137-153. doi:http://dx.doi.org/10.1016/j.nuclphysb.2014.09.013

Li YF (2014) Overview of the Jiangmen Undergorun Neutrino Observatory (JUNO). arXiv:14026143 [hep-ex, Physics:physics] http://arxivorg/abs/14026143

McDonough WF (1990) Constraints on the composition of the continental lithospheric mantle. Earth and Planetary Science Letters 101 (1):1-18. doi:10.1016/0012-821x(90)90119-i

McDonough WF, Sun S-S (1995) The composition of the Earth. Chemical Geology 120:223-253. doi:10.1016/0009-2541(94)00140-4

Reguzzoni M, Sampietro D (2015) GEMMA: An Earth crustal model based on GOCE satellite data. International Journal of Applied Earth Observation and Geoinformation 35, Part A (0):31-43. doi:10.1016/j.jag.2014.04.002

Reguzzoni M, Tselfes N (2009) Optimal multi-step collocation: application to the space-wise approach for GOCE data analysis. Journal of Geodesy 83 (1):13-29. doi:10.1007/s00190-008-0225-x

Shapiro NM, Ritzwoller MH (2002) Monte-Carlo inversion for a global shear-velocity model of the crust and upper mantle. Geophysical Journal International 151:88-105. doi:10.1046/j.1365-246X.2002.01742.x

Strumia A, Vissani F (2003) Precise quasielastic neutrino/nucleon cross-section. Physics Letters B 564 (1-2):42-54. doi:10.1016/s0370-2693(03)00616-6